\documentclass[twocolumn]{aastex6}

\usepackage{epstopdf}
\usepackage{color}

\begin{document}

\title{Star Formation of Merging Disk Galaxies with AGN Feedback Effects}

\author{Jongwon Park$^{1}$, Rory Smith$^{1,2}$ and Sukyoung K. Yi$^{1}$}
\affil{$^1$Department of Astronomy and Yonsei University Observatory, Yonsei University, Seoul 03722, Republic of Korea; jw.park@yonsei.ac.kr\\
$^2$Korea Astronomy and Space Science Institute, Daedeokdae-ro 776, Yuseong-gu, Daejeon 34055, Republic of Korea}

%% Abstract
\begin{abstract}

Using numerical hydrodynamics code, we perform various idealized galaxy merger simulations to study the star formation (SF) of two merging disk galaxies. Our simulations include gas accretion onto supermassive black holes and active galactic nucleus (AGN) feedback. By comparing AGN simulations with those without AGNs, we attempt to understand when the AGN feedback effect is significant. With $\sim$70 simulations, we investigated the SF with the AGN effect in mergers with variety of mass ratios, inclinations, orbits, galaxy structures and morphologies. Using these merger simulations with AGN feedback, we measure merger-driven SF using the burst efficiency parameter introduced by Cox et al. We confirm the previous studies that, in galaxy mergers, AGN suppresses SF more efficiently than in isolated galaxies. However, we additionally find that the effect of AGNs on SF is larger in major mergers than in minor mergers. In minor merger simulations with different primary bulge-to-total ratios, the effect of bulge fraction on the merger-driven SF decreases due to AGN feedback. We create models of Sa, Sb and Sc type galaxies and compare their SF properties while undergoing mergers. With the current AGN prescriptions, the difference in merger-driven SF is not as pronounced as that in the recent observational study of Kaviraj. We discuss the implications of this discrepancy.
\end{abstract}


%%%%%%%%%%%%%%%%%%%%%%%%%%%%%%%%%%%%%%%%%%%%%%%%%%%%%%%%%%%%%%%%%%%%%
%% =============================================================== %%
%% beginning of text ============================================= %%
%% =============================================================== %%
%%%%%%%%%%%%%%%%%%%%%%%%%%%%%%%%%%%%%%%%%%%%%%%%%%%%%%%%%%%%%%%%%%%%%

%%%%%%%%%%%%%%%%%%%%%%%%%%%%%%%%%%%%%%%%%%%%%%%%%%%%%%%%%%%%%%%%%%%%%
%% Section 1 - Introduction %%%%%%%%%%%%%%%%%%%%%%%%%%%%%%%%%%%%%%%%%

\section{Introduction} \label{sec:intro}

According to the $\Lambda$CDM cosmology, galaxy mergers play an essential role in the formation and evolution of galaxies. These events not only change the morphologies of galaxies involved \citep{tt72,netal06}, but also make impressive features such as tidal tails \citep{tt72} or shells \citep{q84} lasting for a long time \citep{jetal14}, or spawn tidal dwarf objects along their tidal features \citep{bh92,bd06,betal07}.

Another important aspect of galaxy mergers is the strong star formation (SF) that occurs during the interactions. \citet{lt78} first suggested that galaxy interactions could trigger strong SF in galaxies \citep{letal84,ketal87,b87}. Observational studies \citep[e.g.,][]{b87} found that a burst of SF occurs in the central regions of interacting galaxies, and \citet{h89}, with his merger simulation, claimed that tidal effect of a companion galaxy triggers a concentration of the gas in the center of the disk, and subsequent merger-driven central SF.

Since then, the understanding of merger-driven SF has continued to broaden with numerical simulations. \citet{mh94} demonstrated that the existence of a bulge in disk galaxies influences the SF driven by minor-mergers and \citet[][hereafter C08]{C08} quantified the amount of merger-driven SF in merging galaxies with various initial conditions. However they did not include the effects of active galactic nucleus (AGN) feedback in their models. 

AGN represent a powerful source of feedback energy. Their location means they can have significant effects on gas that is driven to a galaxy's center during the merging process. The effects of AGN on the SF of merging galaxies was found to be very significant in equal-mass galaxy merger simulations where large quantities of gas are funneled to the galaxy's centers \citep{setal05a,detal05,setal05b}. However, \citet{nk13} found that in isolated galaxies, AGN feedback was less important. \citet{hetal14} performed several merger simulations with AGN feedback, focusing on the comparison between two different types of code, but once again considering only equal mass mergers. 

However, according to \citet{cetal15}, the mass ratio between the two merging galaxies is an important factor that affects the growth of supermassive black holes (SMBHs) and AGN activity. So far a detailed study of the impact of an AGN on SF in non-equal mass mergers has not been conducted, and this forms a central motivation for our study.

%% Figure
\begin{figure*}
\includegraphics[width=0.95\textwidth]{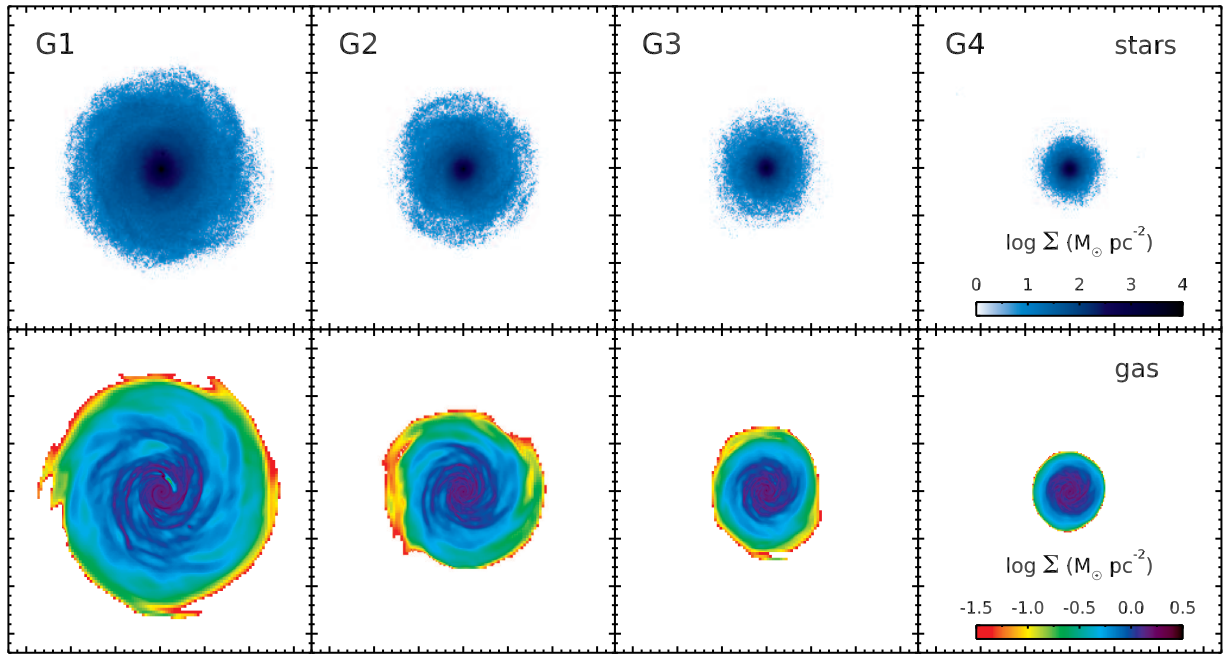}
\caption{Face-on view of four simulated galaxies evolved during 1.0 Gyr. The two-dimensional mass density of the stellar component of simulated galaxies is presented in the top panels and that of the gas component is shown in the bottom panels. The physical scale shown in each panel is 34 kpc.}
\label{fig:SS_iso}
\end{figure*}

The observational study of \citet[][hereafter K14]{k14} demonstrated that the enhancement in the specific star formation rate (SSFR) of disk galaxies experiencing minor mergers is stronger when the galaxy is an Sc or Sd type. K14 argued that this is because the bulge component, which contributes to the stabilization of the disk and suppresses gas inflow and subsequent SF, is small in ``later''-type spiral galaxies. This interpretation was made based on the results of minor merger simulations by \citet{mh94}, but they focussed mainly on the role of the bulge fraction for regulating SF, and did not consider AGN feedback. However, a bulge may suppress gas inflow, which could affect central star formation, but gas inflow can also affect AGN feedback. This AGN feedback, in turn, may affect central star formation. Thus the behaviour of star formation in the presence of a bulge and AGN is highly complex and non-linear, and requires dedicated modelling.

Therefore, in this work, we study the SF of two merging disk galaxies using idealized galaxy merger simulations with AGN feedback, and we will consider a range of mass ratios for both major and minor mergers. Furthermore, we explore the SF caused by minor mergers and its connection with the bulge fraction of primary galaxies with AGN. We quantify merger-driven SF using the burst efficiency defined by C08, for mergers with various initial conditions including type of orbit, inclination, gas fraction and black hole mass, and galaxy morphology.

This paper is organized as follows. In Section \ref{sec:simul}, we explain the simulations we performed, the results of the isolated and merging galaxies are presented in Section \ref{sec:iso} and \ref{sec:merger}, respectively, and we discuss the results in Section \ref{sec:disc}.

% Table - Model
\begin{table*}
  \begin{center}
    \caption{Disk Galaxy Models.}
    \begin{tabular}{lcccc}
    \hline \hline
    \multicolumn{1}{c}{ } & G1 & G2 & G3 & G4 \\
    \hline
    Stellar mass $(M_\odot)$ & $2.00 \times 10^{10}$ & $6.65 \times 10^9$
                                            & $3.33 \times 10^9$     & $2.00 \times 10^9$ \\
                                            
    Stellar disk mass $(M_\odot)$ & $1.60 \times 10^{10}$ & $6.00 \times 10^9$ 
             			                & $3.16 \times 10^9$     & $2.00 \times 10^9$\rule{0pt}{4ex} \\
    Stellar disk scale radius (kpc) & $2.5$ & $2.0$ & $1.5$ & $1.0$ \\
    Stellar disk truncation radius (kpc) & $10.0$ & $8.0$ & $6.0$ & $4.0$ \\
    Stellar disk scale height (kpc) & 0.25 & 0.20 & 0.15 & 0.10 \\
    Stellar disk truncation height (kpc) & 1.0 & 0.8 & 0.6 & 0.4 \\

    Gas disk mass $(M_\odot)$ & $4.00 \times 10^9$ & $2.00 \times 10^9$ 
             	                            & $1.33 \times 10^9$ & $1.00 \times 10^9$\rule{0pt}{4ex} \\
    Gas disk scale radius (kpc) & $4.0$ & $3.2$ & $2.4$ & $1.6$ \\
    Gas disk truncation radius (kpc) & $16.0$ & $12.8$ & $9.6$ & $6.4$ \\
    Gas disk scale height (kpc) & 0.5 & 0.4 & 0.3 & 0.1 \\
    Gas disk truncation height (kpc) & 1.0 & 0.8 & 0.6 & 0.4 \\
    Gas fraction & 0.200 & 0.250 & 0.296 & 0.333 \\

    Bulge mass $(M_\odot)$ & $4.00 \times 10^9$ & $6.65 \times 10^8$ 
             			       & $1.66 \times 10^8$ & $0.00$\rule{0pt}{4ex} \\
    Bulge scale radius (kpc) & $0.5$ & $0.4$ & $0.3$ & $0.2$ \\
    Bulge truncation length (kpc) & $2.5$ & $2.0$ & $1.5$ & $1.0$ \\
    Bulge-to-total ratio & 0.20 & 0.10 & 0.05 & 0.00 \\

    Halo mass $(M_\odot)$ & $6.65 \times 10^{11}$ & $2.22 \times 10^{11}$
                                          & $1.11 \times 10^{11}$ & $6.65 \times 10^{10}$\rule{0pt}{4ex}  \\
    Concentration & 8.40 & 9.47 & 10.22 & 10.81 \\
    $M_*/M_{\rm halo}$ & 0.03 & 0.03 & 0.03 & 0.03 \\

    $M_{\rm BH}$ $(M_\odot)$ & $4.00 \times 10^6$ & $6.65 \times 10^5$
                                               & $1.66 \times 10^5$ & $5.00 \times 10^4$\rule{0pt}{4ex}  \\
    $M_{\rm BH}/M_{\rm bulge}$ & 0.001 & 0.001 & 0.001 & N/A \\

    $N_{\rm total}$ & $3,500,000$ & $1,166,667$ & $583,333$ & $350,000$\rule{0pt}{4ex}  \\
    $N_{\rm disk}$ & $300,000$ & $112,500$ & $59,375$ & $37,500$ \\
    $N_{\rm bulge}$ & $75,000$ & $12,500$ & $3,125$ & $0$ \\
    $N_{\rm halo}$ & $3,125,000$ & $1,041,667$ & $520,833$ & $312,500$ \\

    $m_{\rm disk}$ $(M_\odot)$ & $5.32 \times 10^4$ & $5.32 \times 10^4$
                                                  & $5.32 \times 10^4$ & $5.32 \times 10^4$\rule{0pt}{4ex}  \\
    $m_{\rm bulge}$ $(M_\odot)$ & $5.32 \times 10^4$ & $5.32 \times 10^4$
                                                    & $5.32 \times 10^4$ & $5.32 \times 10^4$ \\
    $m_{\rm halo}$ $(M_\odot)$ & $2.13 \times 10^5$ & $2.13 \times 10^5$
                                                  & $2.13 \times 10^5$ & $2.13 \times 10^5$ \\
                                                  
    Mass ratio with G1 & 1:1 & 3:1 & 6:1 & 10:1\rule{0pt}{4ex}  \\

    \hline
    \end{tabular}
    \label{tab:model}
  \end{center}
\end{table*}

%% Section 1 - Introduction %%%%%%%%%%%%%%%%%%%%%%%%%%%%%%%%%%%%%%%%%
%%%%%%%%%%%%%%%%%%%%%%%%%%%%%%%%%%%%%%%%%%%%%%%%%%%%%%%%%%%%%%%%%%%%%

%%%%%%%%%%%%%%%%%%%%%%%%%%%%%%%%%%%%%%%%%%%%%%%%%%%%%%%%%%%%%%%%%%%%%
%% Section 2 - Simulations %%%%%%%%%%%%%%%%%%%%%%%%%%%%%%%%%%%%%%%%%%

\section{Simulations} \label{sec:simul}

We utilized the adaptive mesh refinement (AMR) hydrodynamics code {\sc ramses} \citep{t02}. This code treats systems of stars or dark matter as collisionless particles and solves the Poisson equation in order to describe their dynamics. Gas follows the Euler equation, and {\sc ramses} determines fluid motions and properties by solving it with the second-order Godunov method.

We choose a  simulation box size of 300 kpc, which is sufficient to contain the galaxies during the merger process. The entire domain is divided using a level 7 coarse grid and the cells can be refined further up to a maximum of level 13 depending on the refinement criteria. This is, if a given cell contains a gas mass greater than $5 \times 10^4 M_\odot$, or the cell size is greater than one-fourth the local Jeans length \citep{tetal97}, it is further refined. The size of the cell with the maximum level of refinement is 300 kpc$/ 2^{13} \approx 37 $ pc.

The SF is modeled based on the Schmidt law, $ \dot{\rho}_{\star} = \epsilon_{\rm sf} \rho_{\rm gas} / t_{\rm ff}$ \citep{s59}, where $\epsilon_{\rm sf}=0.02$ is the star formation efficiency, $\rho_{\rm gas}$ is the gas density of a cell and $t_{\rm ff}=\sqrt{3\pi / 32 G \rho_{\rm gas}}$ is the free-fall time. We choose a SF density threshold of 10.0 H$\ \mbox{cm}^{-3}$, and the corresponding masses of new-born star particles are $1.58 \times 10^4 M_\odot$. The number of new star particles are determined by the Poisson distribution $P(N)=\lambda^N/N! \ \exp(-\lambda)$, where $\lambda = \epsilon_{\rm sf} \ (\rho_{gas} \Delta x^3 / m_*) \ (\Delta t / t_{\rm ff})$ is the mean and $\Delta x$ is the size of the star forming cell. We adopted the kinetic supernova feedback of \citet{dt08}. We assume that the fraction of mass that evolves to the supernovae in each star particle $\eta_{\rm SN}$ is 0.1, the mass loading factor $\eta_{\rm W}$ is 1.0, and the fraction of supernovae energy released in the kinetic form is 0.5. The supernovae bubble radius is 75 pc.

We adopted the equation of state of \citet{betal10}. In this approach, the interstellar medium is modeled assuming equilibrium between cooling (atomic and molecular) and heating by ultraviolet radiation. This helps to save time for the calculation of those processes during the simulation. The equation of state is the following; for a density of $10^{-3} < n < 0.3 \ {\rm H\ cm^{-3}}$, $T=10^4 {\rm K}$. Below  $10^{-3} \ {\rm H\ cm^{-3}}$, $T=4 \times 10^6 \ (n/10^{-3})^{2/3}$ and $T=10^4 \ (n/0.3)^{-1/2}$ K above $0.3 \ {\rm H\ cm^{-3}}$. We force the grid refinement to ensure that the thermal Jeans length of the gas is always resolved by at least four cells \citep{tetal97}. However, in the densest gas, this may surpass our maximum refinement level. Thus, we also use a temperature floor for the highest density gas, which can be considered as a subgrid model for the unresolved turbulent motions of the gas at scales smaller than our maximum grid resolution \citep{betal10,tetal10}.

In {\sc ramses}, the gas accretion to SMBHs and AGN feedback are implemented. This has been accomplished in a variety of ways in the past \citep[e.g.][]{detal09,tetal11,detal12}. However, we follow the AGN prescription of \citet{gb13} who simulated SMBHs in disk galaxies. SMBHs accrete gas around it at the Bondi-Hoyle accretion rate \citep{bh44,b52}

\begin{equation} \label{eq:acc}
\dot{M}_{\rm BH} = \alpha \frac{4 \pi G^2 M^2_{\rm BH} \rho}{(c^2_s + u^2)^{3/2}}
\label{eq:bondi-hoyle}
\end{equation}
where $M_{\rm BH}$ is the black hole (BH) mass, $G$ is the gravitational constant, $\rho$ is the gas density, $c_s$ is the gas sound speed, $u$ is the gas velocity relative to the SMBH, and $\alpha$ is the boost factor. For the computation of the gas accretion rate, gas cells within the radius of $4 \Delta x$ are considered. Here, $\Delta x$ is the size of the cell with the maximum refinement level. The average gas properties are calculated using the properties of the individual cells within this radius, weighted by the distance from the black hole \citep{ketal04}. Larger weights are given to the cells close to BH. In cosmological simulations, it is assumed that the boost factor is greater than unity \citep{tetal11,detal12}, because the limitation of spatial resolution and the difficulty of modeling of cold interstellar medium causes an underestimation of accretion rate \citep{bs09}. In this work, we set $\alpha=1$ as the resolution of our simulations are higher than those of such cosmological simulations and we can model the cold interstellar medium thanks to the equation of state model of \citet{betal10}. We performed an additional equal-mass merger simulation with density-dependent $\alpha$ following \citet{bs09} and found that our choice of boost factor does not affect the integrated SF significantly. The gas accretion rate is limited by the Eddington limit.
\begin{equation} \label{eq:Edd}
\dot{M}_{\rm Edd} = \frac{4 \pi G M_{\rm BH} m_{\rm p}}{\epsilon_{\rm r} \sigma_{\rm T} c}
\label{eq:Edd_limit}
\end{equation}
Here, $m_{\rm p}$ is the proton mass, $\sigma_{\rm T}$ is the Thomson cross-section, $c$ is the speed of light, and $\epsilon_{\rm r}$ is the efficiency by which accreted gas mass is converted into luminous energy. At each coarse time step, thermal energy is injected into the ambient gas (quasar mode). The amount of energy is

\begin{equation} \label{eq:FB}
\Delta E_{\rm acc} = \epsilon_{\rm c} \epsilon_{\rm r} \dot{M}_{\rm acc} c^2 dt
\label{eq:AGNFB}
\end{equation}
The value of the coupling efficiency $\epsilon_{\rm c}$ is 0.15 \citep{bs09,tetal11}. This energy injection occurs when the thermal energy can increase the weighted averaged temperature of the cells to a minimum temperature $T_{\rm min}=10^7 ~{\rm K}$. In \citet{gb13}, there is a maximum temperature $T_{\rm max}=5 \times 10^9 ~{\rm K}$ which prevents excess heating, but we confirm it rarely occurs in our merger simulations.

\begin{figure}
\includegraphics[width=0.45\textwidth]{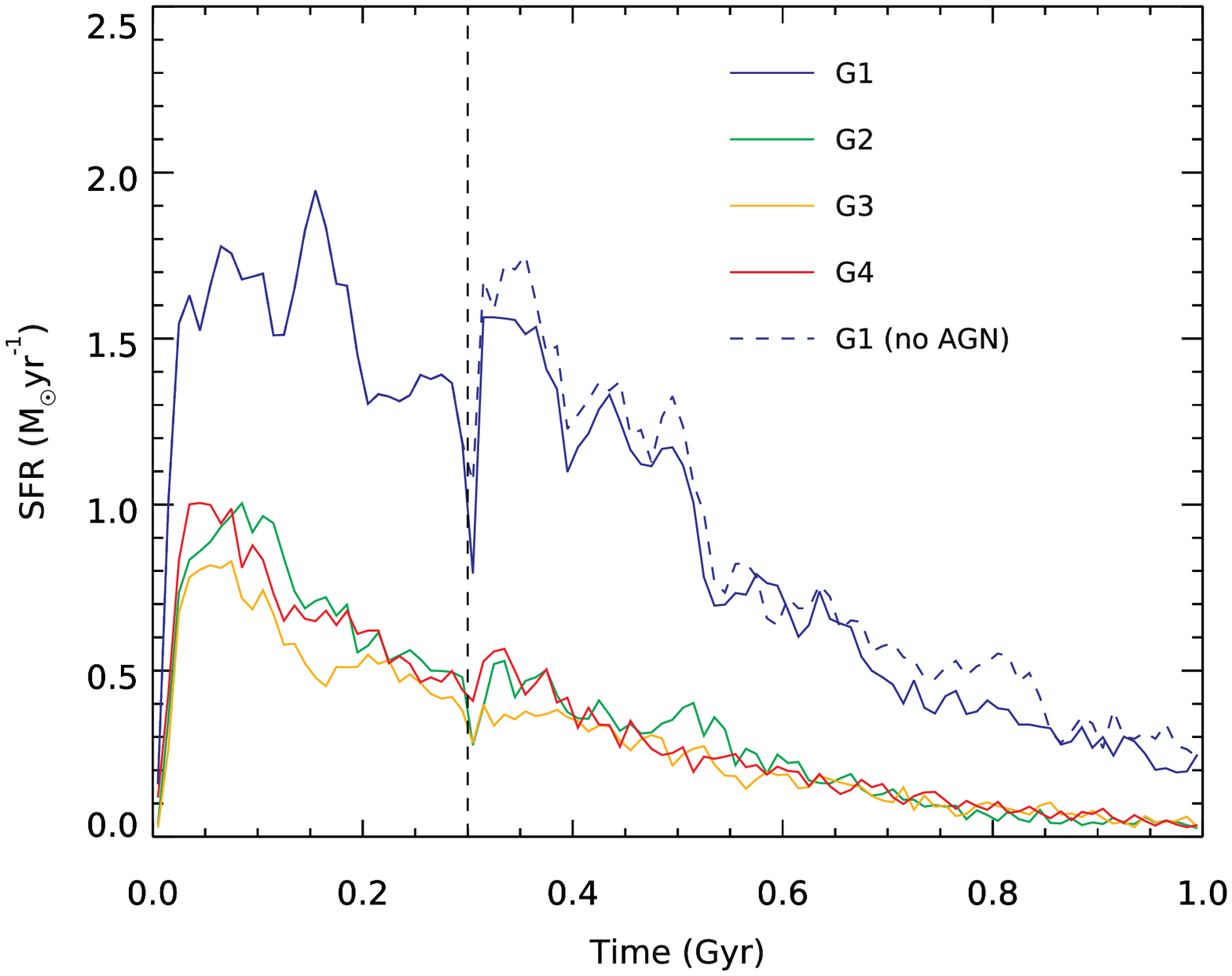}
\caption{SFHs of galaxies in isolation. Each colored solid line means the SFH of the galaxy with AGN feedback. We also present the SFR of isolated G1 without AGN feedback (dashed blue line). The vertical dashed line indicates 0.3 Gyr. After this time, the spatial resolution increases from level 11 to 13 with the SF density threshold from 0.1 to 10.0  H${\rm \ cm}^{-3}$. There are sudden increases in SFR when the resolution changes.}
\label{fig:SFH_1234}
\end{figure}

%% Section 2 - Simulations %%%%%%%%%%%%%%%%%%%%%%%%%%%%%%%%%%%%%%%%%%
%%%%%%%%%%%%%%%%%%%%%%%%%%%%%%%%%%%%%%%%%%%%%%%%%%%%%%%%%%%%%%%%%%%%%

%%%%%%%%%%%%%%%%%%%%%%%%%%%%%%%%%%%%%%%%%%%%%%%%%%%%%%%%%%%%%%%%%%%%%
%% Section 3 - Isolated Galaxies %%%%%%%%%%%%%%%%%%%%%%%%%%%%%%%%%%%%

\section{Isolated Galaxies} \label{sec:iso}

\subsection{Initial Conditions}

Our disk galaxy models consist of a dark matter halo, stellar disk, gas disk and stellar bulge. The density profile of the dark matter halo is described by an NFW profile \citep{NFW96}. We choose values for the halo concentration that are consistent with $\Lambda$ cold dark matter cosmological simulations \citep{netal07,petal12,correa15,ketal16}, and follow the usual $z=0$ trend of increasing concentration with decreasing mass. The ratio of stellar mass to dark matter halo mass depends on the halo mass \citep{metal10} and it may differ even if two dark matter haloes have the same mass \citep{fetal12,metal14}. However, we decide to fix this ratio to 0.03, because changing stellar mass leads to the change of other related parameters such as gas fraction or bulge size and makes our experiments more complex to interpret. The stellar disk follows an exponential profile radially and $sech^2\,z$ profile vertically. We determine the radial scale lengths of stellar disks based on the r-band scale length-stellar mass relation seen in observed galaxies \citep{fetal10}. The truncation radius of the stellar disks is four times the scale radius. For example, G1, has a scale radius of 2.5 kpc and a truncation radius of 10 kpc. The vertical scale height and truncation height are initially 10\% of the radial scale lengths (e.g. 0.25 and 1.0 kpc for G1). The gas disk follows an exponential profile both radially and vertically, and its radii (both scale and truncation) are $1.6$ times those of the stellar disk \citep{cetal94}. The vertical scale height of the gas disks are given in Table \ref{tab:model}. The amount of gas in a galaxy is determined by its stellar mass \citep{getal08,cetal11}. The density profile of the stellar bulge component follows the profile of \citet{h90}.

Our galaxies are modeled to mimic local spiral galaxies from Sb to Sd type. The bulge to total ratio (B/T, bulge mass divided by the sum of bulge and stellar disk mass) decreases from large to small galaxies. We determine the values following the B/T light ratios of galaxies with different morphologies \citep{gw08}. We model our largest galaxy, G1, after an Sb-like galaxy  because they are abundant in the local Universe and mergers including Sb primary galaxies are more common \citep{ketal15}. We also produce Sa- or Sc-like primary galaxies and their merger-driven SF properties are discussed in Section \ref{ssec:morphology}. The smaller galaxies become increasingly late-type as we move from G2, to G3, and to G4. 

We fixed the $M_{\rm BH}/M_{\rm bulge}$ for all galaxies except G4, which is bulgeless. We confirm that this choice is allowed within the uncertainties in the relation between the BH mass and bulge mass of observed galaxies \citep{mh03,betal11}.

The face-on images of the simulated disk galaxies, evolved in isolation for 1.0 Gyr, are presented in Figure \ref{fig:SS_iso}, and the details of the disk galaxy models are provided in Table \ref{tab:model}.

\subsection{Evolution of Isolated Galaxies} \label{ssec:isoevo}

We performed simulations with low resolution (level 11) for the first 0.3 Gyr, which is roughly similar to the dynamical time scale of G1. This is because all the components of a disk galaxy are not in equilibrium and, hence, it causes density perturbations. We allow SF to occur with a density threshold of $0.1 ~{\rm H \ cm^{-3}}$, but we did not allow the gas accretion and AGN feedback to occur. During this period, a ring-like structure appears in the disk, moves outwards from the center to the outskirts and then disappears. After 0.3 Gyr, we run the simulations with a level 13 resolution and the SF density threshold is $10.0 ~{\rm H \ cm^{-3}}$. Gas accretion and AGN feedback are switched on after this time. Numerical effects resulting from the change in resolution are very short-lived because cooling timescales are significantly shorter than dynamical timescales. Figure \ref{fig:SFH_1234} presents the star formation histories (SFHs) of galaxies evolving in an isolated environment. The largest galaxy, G1, exhibits the greatest SFR (blue lines) and the SFRs of the other three galaxies (G2, G3 and G4) are similar to each other (see the green, orange and red line). As AGN feedback is not allowed to occur during the first 0.3 Gyr, the AGN feedback affects the SFR only after 0.3 Gyr (blue solid and dashed line). 

We confirm the results of \citet{nk13}, that for isolated galaxies AGN does not strongly influence the star formation. For example, in our G1 model the star formation over 6 Gyrs is only reduced by 13.2$\%$ by the presence of an AGN.
 
%% Section 3 - Isolated Galaxies %%%%%%%%%%%%%%%%%%%%%%%%%%%%%%%%%%%%
%%%%%%%%%%%%%%%%%%%%%%%%%%%%%%%%%%%%%%%%%%%%%%%%%%%%%%%%%%%%%%%%%%%%%

%%%%%%%%%%%%%%%%%%%%%%%%%%%%%%%%%%%%%%%%%%%%%%%%%%%%%%%%%%%%%%%%%%%%%
%% Section 4 - Mergers %%%%%%%%%%%%%%%%%%%%%%%%%%%%%%%%%%%%%%%%%%%%%%

\section{Mergers} \label{sec:merger}

In this section, we discuss the SF properties of various merging galaxies. For clarity, we define terms that will be frequently used throughout this paper. In non-equal merger cases, we call the larger galaxy and smaller galaxy the primary galaxy and the secondary galaxy, respectively. We define the merger mass ratio as the ratio of the stellar mass of the primary galaxy to the stellar mass of the secondary galaxy, that is, $M_{*, \rm primary}/M_{*, \rm secondary}$. We deal with galaxy mergers with mass ratios of 1:1, 3:1 (major), 6:1 and 10:1 (minor).

\begin{figure*}
\includegraphics[width=0.95\textwidth]{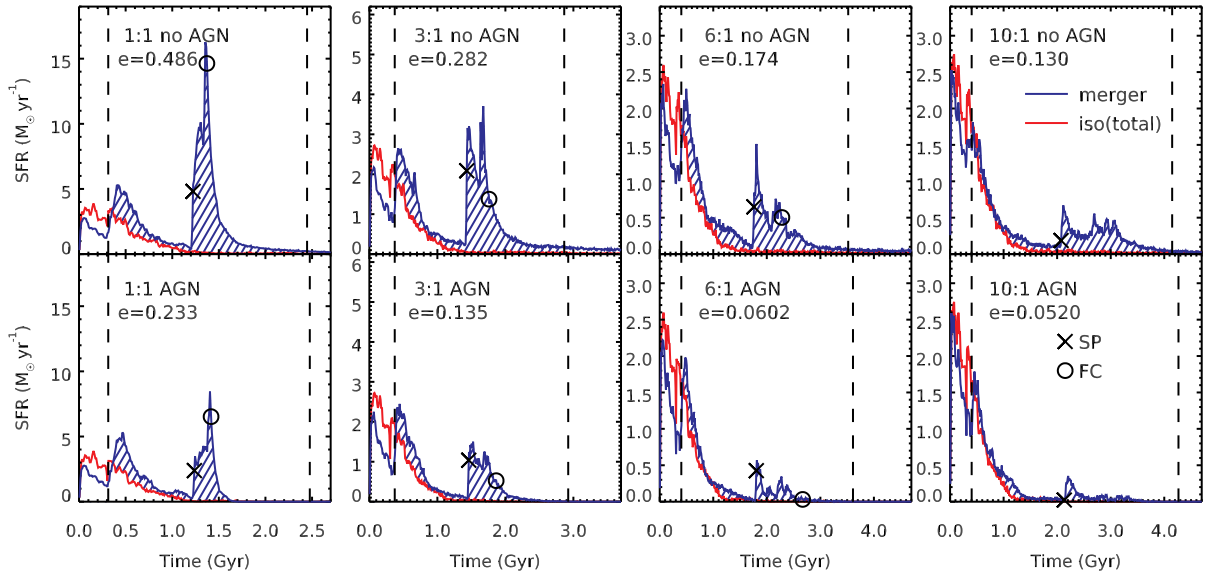}
\caption{SFHs of merging galaxies with different mass ratios. The upper panels present the results of non-AGN simulations while the lower panels show those of AGN simulations. The mass ratios are 1:1, 3:1, 6:1 and 10:1 from left to right. In each panel, the red line shows the sum of the SFRs of isolated galaxies and the blue line indicates the total SFR of the merging system. The shaded region between those two lines means the merger-driven SF that occurs between FP and 2$\times$SP (shown by vertical dashed lines). SP and FC are represented with symbols (see the legend). The burst efficiency ($e$) is given in each panel. The orbit is parabolic for all mergers.}
\label{fig:SFH_mass}
\end{figure*}

Each merger simulation starts with two galaxies separated by an initial distance of $0.8 R_{vir}$, where $R_{vir}$ is the virial radius of a primary galaxy. The initial positions and velocities of the galaxies are determined by the orbits we want to simulate (elliptical, parabolic or hyperbolic) with the center of mass of the two galaxies located at the simulation box center. However, in 10:1 mergers with hyperbolic orbits, we shifted the center of mass from the box center to prevent secondary galaxies with large initial velocities from escaping from the box after the first encounter of two galaxies.

When the separation between the two galaxies becomes a minimum for the first time, we call it the first passage (FP). The second minimum separation is defined as the second passage (SP). We define the final coalescence (FC) by the moment when the centers of the two galaxies are closer than 1 kpc \citep{letal08}. We use the center of mass of the bulge component as the galaxy center, as it is small and concentrated, except in the 10:1 mergers where the lower mass galaxy lacks a bulge. Then we use the center of mass of the stellar disk instead.

As with the isolated galaxy simulations discussed in Section ~\ref{ssec:isoevo}, the merger simulations are initially performed with a lower resolution for the first 0.3 Gyr, to ensure stability \citep[similar to in][]{getal16}, before swtiching to the high resolution phase. During this period, gas accretion and AGN feedback are switched off. We confirm that FP and the corresponding burst of SF always occur during the high resolution phase.

In any merger, there are a vast number of parameters that can be varied. In order to conduct our parameter study, we follow the following procedure. In order to do so, we fix all other parameters, while varying only a single parameter, so as to clearly see the dependency of the SF on the parameter (Section \ref{ssec:inc} and \ref{ssec:orb}). 

\subsection{Burst Efficiency} \label{ssec:burst}

\begin{figure*}
\includegraphics[width=0.95\textwidth]{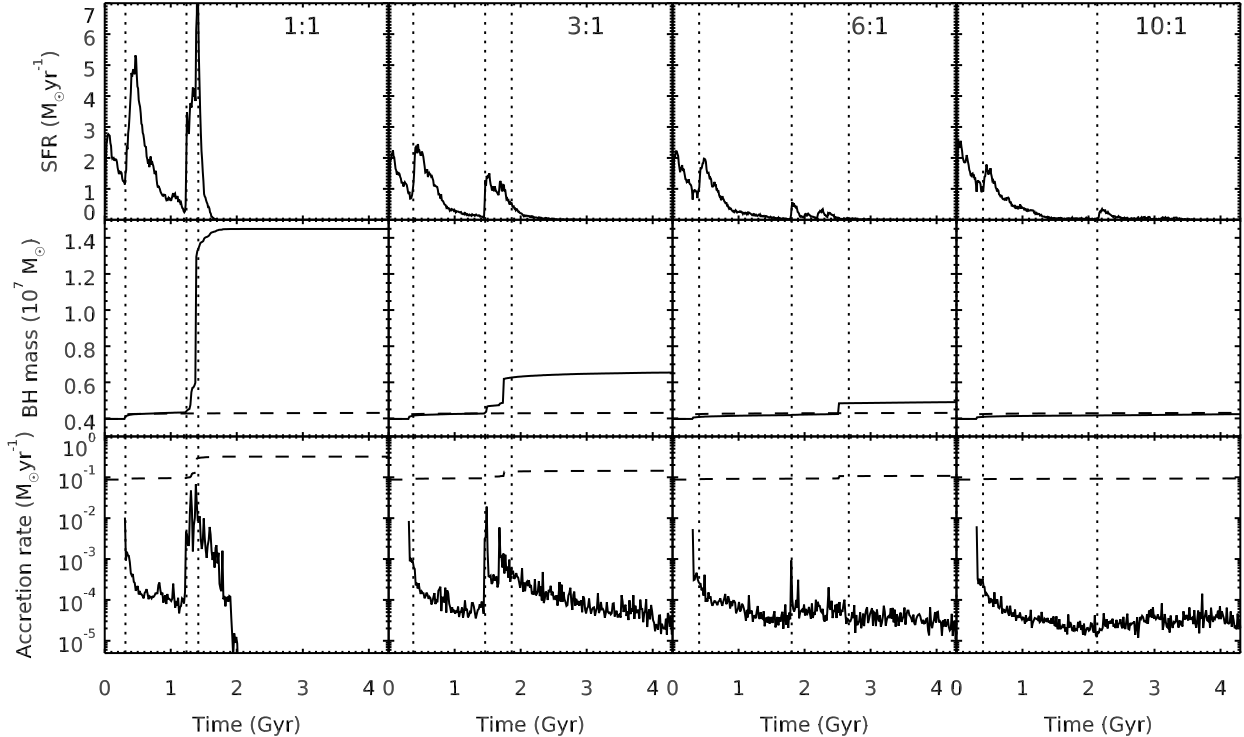}
\caption{BH growths in AGN simulations with mass ratios 1:1, 3:1, 6:1 and 10:1. The top panels present the evolution of the SFRs which are the same as those in the lower panel of Figure \ref{fig:SFH_mass}. Panels in the middle row displays the mass of the SMBH of the primary galaxy. For comparison, the BH mass of the isolated G1 model is indicated by a dashed line. The bottom panels show the accretion rates of primary BHs with Eddington limits. FP, SP and FC time are given with vertical dotted lines. Note that the AGN starts at 0.3 Gyr (Section \ref{ssec:isoevo}).}
\label{fig:acc}
\end{figure*}

In order to quantify the merger-driven SF, we used burst efficiency $e$ which was used by C08. This parameter is defined by the difference between the fraction of gas consumed by SF in the interacting system and that in isolation. If the burst efficiency of a galaxy merger is zero, it means that the merger does not trigger the starburst at all and the SFHs of isolated and merging galaxies are identical. If this value is 0.5, it means that 50\% of the initial gas mass has been converted to merger-driven SF. When measuring the burst efficiency, we consider the SF that occurs between FP and twice the SP. The reason for starting at FP and finishing at twice SP is motivated below.

Firstly, when calculating the burst efficiency, we only consider SF that occurs after FP. This is important as it helps to mitigate numerical effects during the initial phase of the simulation when resolution is low. During this early phase, we find a numerical effect where the SFR of a galaxy that is moving face-on is reduced by its motion through the grid. This effect is partly responsible for difference between SFR of the merging galaxies and their isolated counterparts that can be seen during the low resolution, early phase of our simulations. However, we expect that its influence on the later evolution, when the resolution is higher, is much more limited. This is because, in order to keep the combined center of mass of the merging galaxies near the center of simulation box in a minor merger, the primary galaxy, which contains the majority of the gas, actually moves slowly. Also the bursts of SF occur after dynamical friction has acted to significantly slow down the motion of the galaxies in order that they can merge.

Next, we consider our choice of using twice the SP as the end point for measuring the burst efficiency. C08 measured the burst efficiency by considering SF that takes place during the period of 6.0 Gyr for all their simulations. However, we vary this period for each simulation because each merger ends at a different time. \citet{letal08} defined the post-merger as FC + 1 Gyr. Instead of using either of these definitions, we measure the SF between FP and twice the SP. This is partly because the absence of the stellar bulge in G4 makes it difficult to determine the FC in 10:1 mergers. However, in practice we find that the two time scales (FC + 1 Gyr or $2 ~\times$ SP) match well and, in any case, the burst efficiencies are very similar to each other. For instance, in 1:1 merger simulation without AGN feedback, the SP time scale is 1.22 Gyr and the FC time scale is 1.37 Gyr. The burst efficiency measured with the SP time and with the FC time is 0.486 and 0.485, respectively. The reason why they are so similar is because there is very little star formation towards the end of the simulations and so, although we choose a factor of 2 arbitrarily, our results are not sensitive to this choice.

\subsection{AGN Feedback and Mass Ratios} \label{ssec:mass}

\begin{figure*}
\includegraphics[width=0.95\textwidth]{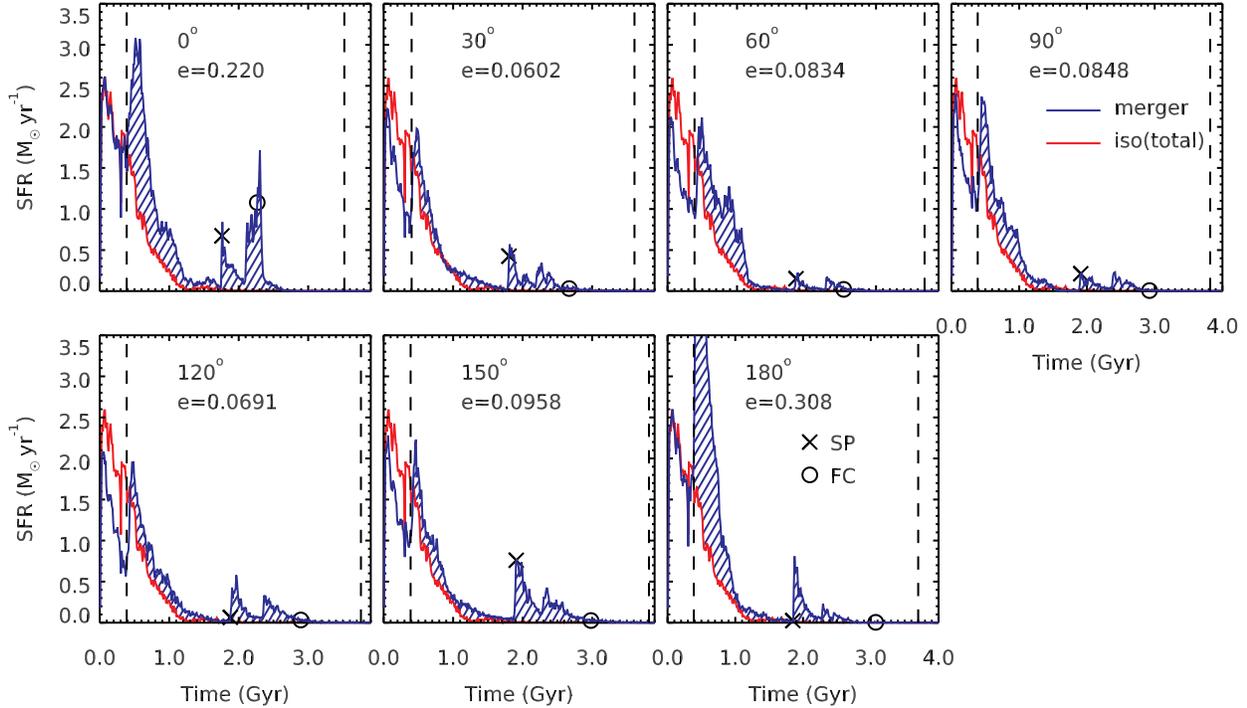}
\caption{SFHs of 6:1 mergers with various inclinations. The format is the same as in Figure \ref{fig:SFH_mass}. The inclination angle and burst efficiency are given in each panel. The orbit is parabolic and AGN feedback is considered in all merger simulations presented in this figure.}
\label{fig:SFH_inclination}
\end{figure*}

In the studies of the effect of AGN feedback on SF in galaxy mergers \citep{detal05,nk13,hetal14}, only equal-mass mergers were targeted and simulated. However, a suite of merger simulations by \citet{cetal15} demonstrated that the mass ratio of two galaxies is the most important factor that determines the growth of SMBHs and AGN activity. In order to see the effect the AGN on SF in mergers with different mass ratios, we compare merger simulations (with and without AGN feedback) for four mass ratios (1:1, 3:1, 6:1 and 10:1). These are created with the combination of G1 and one of the four galaxies from G1 to G4.

Figure \ref{fig:SFH_mass} presents the SFHs of these simulations. A decrease in SFR by AGN feedback is observed in the mergers of all mass ratios. Independent of mass ratio, the SF shortly after FP is not affected significantly by the AGN feedback but the decrease in SF after SP is more pronounced. This is because the concentration of gas in the galaxy center after FP is not as intense as the concentration at SP or FC. Immediately after FP, SF still occurs throughout the disks, thus, the AGN effect is small. However, at a late evolutionary stage such as FC, bursts of SF are severely affected by AGN activity because the gas is concentrated in the galaxy center, surrounding the AGN.

Not only the instantaneous SFR but also integrated SF is significantly affected. We measure the total mass of new stars that form during the first 6 Gyr for our simulations. In isolation, our G1 model produces on 13.2\% less mass in stars, when an AGN is included. But in an equal-mass merger of two G1 galaxies, 40.4\% less stellar mass is produced by including the effects of AGN feedback. This is consistent with the findings of \citet{nk13} which indicates that AGNs have a small impact in isolated galaxies while the suppression of SF by AGNs is more dominant in galaxy mergers, which tend to drive gas to the galaxy centers, feeding the AGN. We find that in 3:1, 6:1 and 10:1 mergers, the total stellar mass produced in 6~Gyr is reduced by 39.5, 37.8 and 30.5\%, respectively, indicating that the more major mergers are better able to drive gas onto the AGN.

\begin{figure*}
\includegraphics[width=0.95\textwidth]{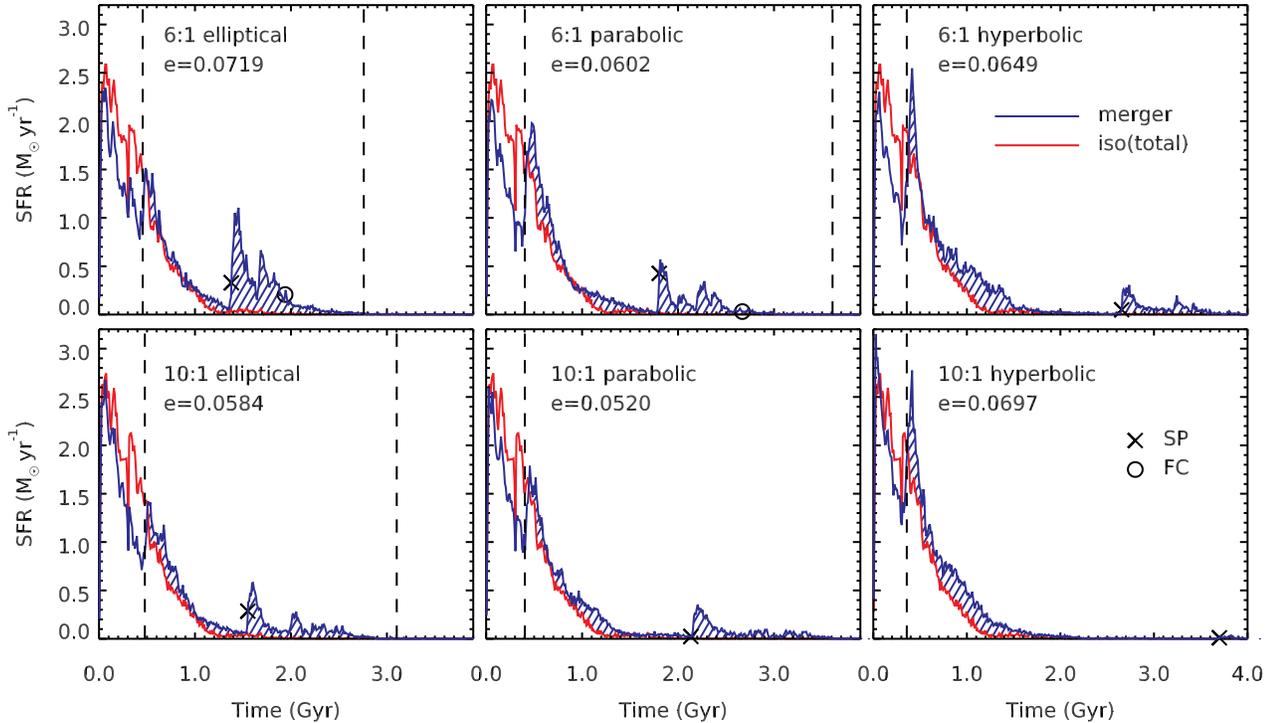}
\caption{SFHs of minor mergers with three different orbits. In each panel, mass ratios and orbits and burst efficiency are given. AGN feedback is considered.}
\label{fig:SFH_orbit}
\end{figure*}

For equal mass ratio mergers, the burst efficiency is 0.486 without AGNs, and 0.233 with AGN feedback, corresponding to 52\% of its original value. The 3:1 merger shows a similar proportion of decrease in the value of burst efficiency (from 0.282 to 0.135, 52\% reduction). The minor mergers exhibit an even greater decrease in burst efficiencies (65\% and 60\%, respectively for 6:1 and 10:1 mergers).

Difference in the mass growth of SMBHs accounts for the variations in the extent of SF suppression by AGN feedback in mergers with different mass ratios. In Figure \ref{fig:acc}, we present the SMBH activities of merging galaxies. In the equal-mass merger, the mass growth of the BH is modest at FP and SP. But at FC, both the BH merger and intense gas accretion contribute to a large growth of the BH. For increasingly minor mergers, the mass growth of the SMBH becomes less pronounced and the amount of gas accretion decreases. For our most minor merger (10:1 mass ratio) the SMBH does not show any enhancement in BH activity. This result is in agreement with \citet{cetal15}, who claimed that major mergers have a significant impact on the growth of the primary galaxy's SMBH due to the strong tidal torques by large companion galaxies. The moment when the highest accretion rate occurs also changes depending on merger mass ratio. For example, the highest accretion rate appears before FC in the 1:1 merger while the accretion rate is highest before SP in the 3:1 and 6:1 merger. The duration that accretion rates are maintained also depends on mass ratio. In equal-mass mergers, the accretion rate becomes almost zero after 2 Gyr, while the SMBHs in the other mergers maintain their gas accretion.

\subsection{Inclination} \label{ssec:inc}

\begin{figure*}
\includegraphics[width=0.95\textwidth]{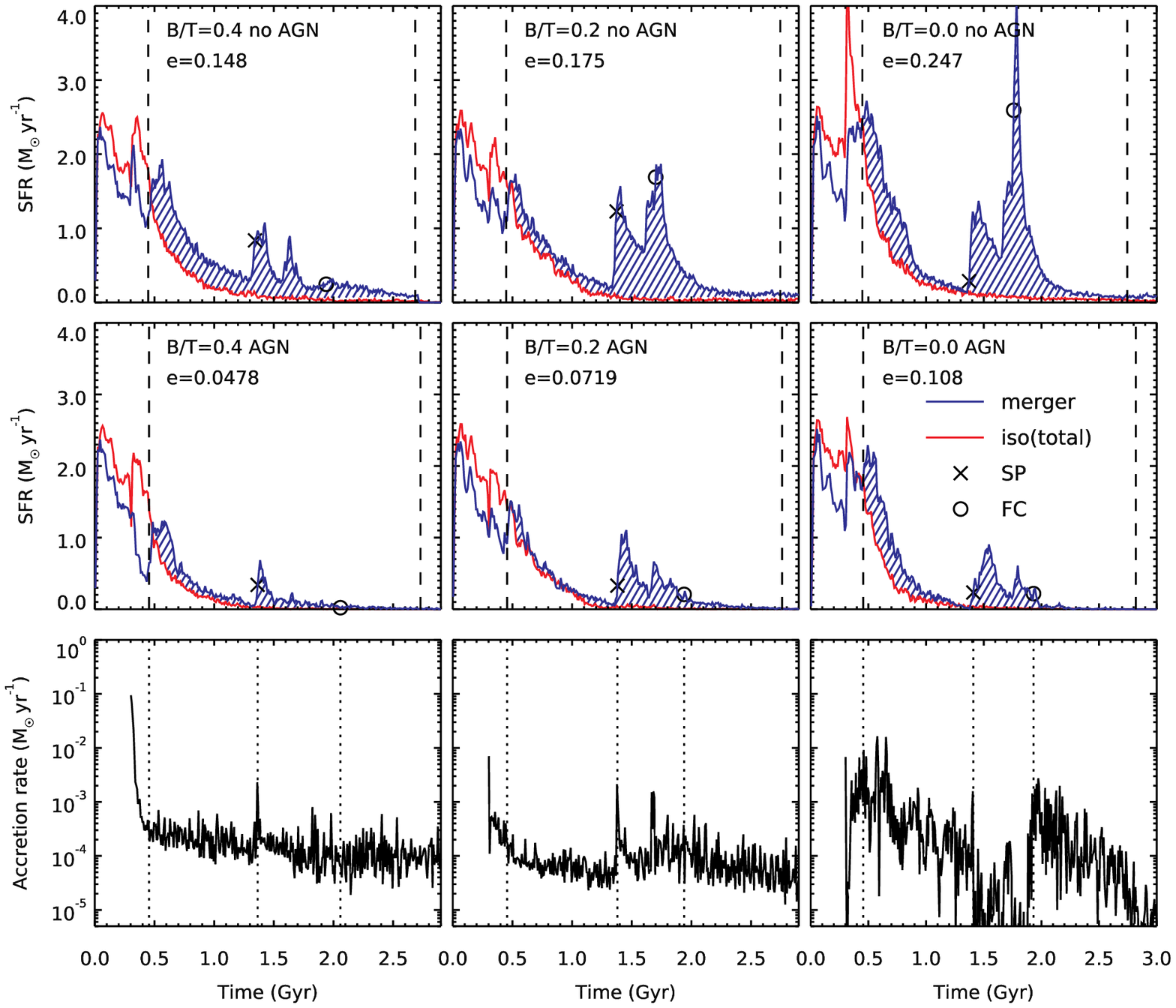}
\caption{SF and SMBH activities of merging galaxies with various B/T ratios. The top panels show the SFRs of simulations without AGN feedback and the middle panels show those of the simulations with the AGN. The accretion rate of the primary SMBH of the AGN simulations is given in the bottom three panels with the same format as used in Figure \ref{fig:acc}.}
\label{fig:SFH_bulge}
\end{figure*}

We measured the burst efficiencies of mergers between G1 and G3 (mass ratio 6:1) with AGN feedback by changing the angle between the spin plane of the primary galaxy (G1) and orbital plane of the secondary galaxy (G3). We fix the galactic disk of the secondary galaxy to be in its orbital plane.

As mentioned in Section \ref{ssec:burst}, the inclination of a galaxy can lead to an artificially lowered rate of SF due to numerical effect. This can be seen in difference between the star formation rates between the isolated and merger model during the low resolution phase (before 0.3 Gyr) in Figure \ref{fig:SFH_inclination}. We minimize the impact of this effect by neglecting the contribution of star formation during the lower resolution phase, when numerical effects will be most strong. We find that inclination angles of coplaner mergers ($0^\circ$ and $180^\circ$ inclination) have larger increases in SFRs. This effect must be partly physical, as the merger simulations of C08 were conducted with an SPH code which does not suffer from numerical diffusion, and found similar results with regards to the fact that coplanar mergers result in the highest star formation enhancement.

Although prograde ($0^\circ$) and retrograde ($180^\circ$) both show larger SFRs overall, the time at which the greatest enhancement of star formation occurs is different. In prograde mergers it occurs at FC, meanwhile in retrograde mergers it occurs at FP.

In C08, the total SF over the whole simulation period was reported to be higher in prograde merger, in contrast to \citet{detal07} who reported more efficient SF in retrograde mergers. Our results shows a larger value of the burst efficiency in the retrograde merger, driven primarily by the strength of the starburst that occurs immediately after FP.

\subsection{Orbits} \label{ssec:orb}

Orbits are determined by the initial positions and velocities of secondary galaxies. In Figure \ref{fig:SFH_orbit}, we present the SFHs of minor mergers (mass ratio 6:1 and 10:1) with different orbits. We consider three different types of orbits. For a given mass ratio, the secondary galaxy with an elliptical orbit (eccentricity $\epsilon\footnote{in this work $\epsilon$ denotes eccentricity because $e$ is used for burst efficiency}=0.97$) possesses the lowest kinetic energy of all the orbits. The initial kinetic energy increases as the eccentricity of the orbit increases. The parabolic orbit has $\epsilon=1.00$, and the hyperbolic orbit has $\epsilon=1.03$. Our choice of a maximum eccentricity of 1.03 is because if it is larger then in the 10:1 hyberbolic merger the secondary galaxy comes too close to the simulation box boundaries.

For a given mass ratio, SF peak after the FP is highest in a hyperbolic orbit. In this orbit, the secondary galaxy moves faster and encounter the primary galaxy earlier than that in the other orbits. There is more gas available at this early moment leading to a higher SFR. On the other hand, it takes longer time for the secondary galaxy to finally merge with the primary galaxy on such an energetic orbit. Since less gas remains at this late epoch, SFR after the SP is lower in such an orbit. The opposite is true in an elliptical orbit: the SF peak is lowest after the FP and highest after the SP, compared to its counterparts on other orbits. Because of this, there is no simple correlation between the shape of the orbit and the burst efficiency observed.

\subsection{Bulge Fractions} \label{ssec:bulge}

Using observational data, K14 demonstrated that the enhancement in the SSFR of spiral galaxies possessing disturbed features (which are thought to be caused by minor mergers) is more prominent in later-type spiral galaxies such as Sc and Sd types. He interpreted that this is because those galaxies have low bulge-to-total ratio ($M_{\rm bulge}/(M_{\rm bulge}+M_{\rm stellar ~disk})$) and high gas fraction ($M_{\rm gas ~disk}/(M_{\rm gas ~disk}+M_{\rm stellar ~disk})$). In this section, we consider the effect of the bulge on the merger-driven SF, and fix the gas fraction. The effect of the gas fraction will be treated in Section \ref{ssec:gas}.

The interpretation of K14 about bulge fraction and SF enhancement is based on the minor merger simulations of \citet{mh94}. In their study, the bulgeless disk galaxy shows larger SFR enhancement compared to the one with a bulge component. They suggested that gas inflow and the subsequent starburst are regulated when there is a central bulge. The simulations of C08 also varied the bulge-to-total ratio in the case of minor mergers, and found higher values of burst efficiency for lower bulge-to-total ratio. However, both of these previous studies lacked a treatment of AGN feedback.

Similar to \citet{mh94}, we fix the stellar disk mass while varying the B/T ratio with values of 0.0, 0.2, and 0.4. This is conducted on the primary galaxy, which has the properties of G1 (except of course the bulge). The secondary galaxy is always G3.

\begin{figure*}
\includegraphics[width=0.95\textwidth]{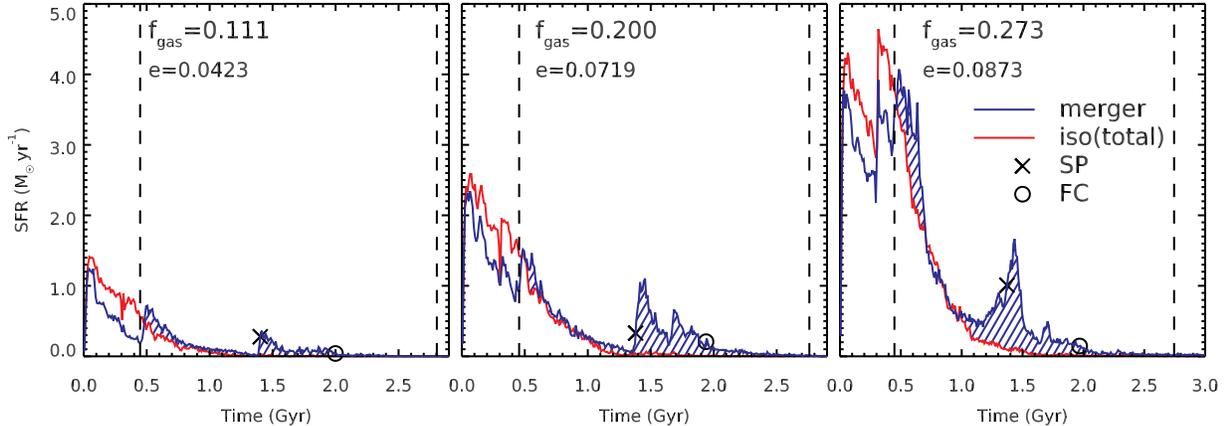}
\caption{SFRs of merging galaxies with different gas fractions. The assumed mass ratio is 6:1 and an elliptical orbit is adopted. The stellar mass is fixed but only the gas mass varies.}
\label{fig:SFH_gas}
\end{figure*}

Figure \ref{fig:SFH_bulge} shows the resulting SFHs and accretion rates of these simulations. When the AGN feedback effect is ignored (top three panels), the bulgeless galaxy exhibits a larger instantaneous SFR at FC and burst efficiency than those of the others. However, these differences are less prominent when AGN feedback is considered (middle panels). This is because the central concentration of the gas, which is less suppressed in low B/T galaxies, triggers not only the starburst but also the intense gas accretion and subsequent AGN activity. In the bottom panels, it is shown that the gas accretion rate in a bulgeless galaxy (right panel) is maintained at a higher value than in the other galaxies between FP and SP. BH activity becomes weaker after SP but it increases again after FC.

With AGN feedback, the burst efficiency still increases with a decreasing B/T ratio. However, the peak instantaneous SFR, espcially around FC, does not show such a clear trend with B/T ratio. This suggests that the dependence of merger-driven SF on the bulge fraction seen in previous studies is weakened when AGN feedback is included.

\subsection{Gas Fraction} \label{ssec:gas}
Another factor suggested by K14 that causes a stronger enhancement in SSFR in late-spirals is gas fraction. We present the SFHs of 6:1 mergers (merger between primary galaxy and G3) with different primary galaxy gas fractions in Figure \ref{fig:SFH_gas}. The primary galaxy with $f_{\rm gas}=0.200$ is G1, but we vary the gas fraction while fixing the stellar mass.

\begin{figure*}
\includegraphics[width=0.95\textwidth]{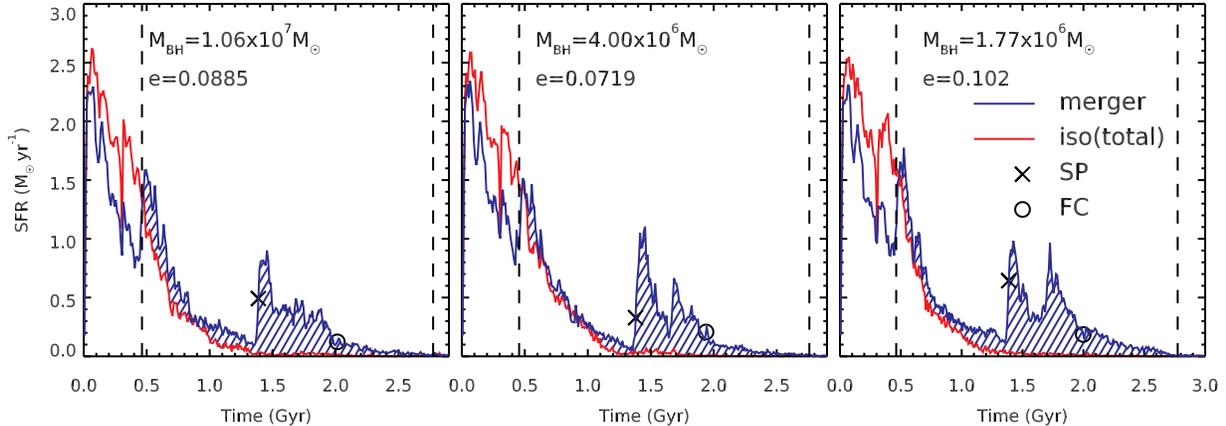}
\caption{SFHs of mergers with BH masses from high to low (from left to right). The mass ratio of 6:1 and an elliptical orbit are assumed. The primary galaxies in the three simulations have identical bulge masses.}
\label{fig:SFH_BH}
\end{figure*}

In this suite of simulations, mergers with high gas fractions show higher burst efficiencies. This is opposite to the result of C08. They claimed that galaxies with a high gas fraction consume a large amount of gas regardless of a merger, therefore, those galaxies show smaller value of burst efficiency. In our simulations, it is true that higher gas fractions do lead to higher SFRs in isolated galaxies. However, regardless of this, our mergers with higher gas fractions do show increased SFRs, suggesting that our galaxies still have sufficeint gas for a starburst when the mergers occur. The reason for this difference may be two fold. Perhaps their mergers occur later, when more of the gas has been used up. Or perhaps their sub-grid physics treatment of star formation differs from our own.

The increase in SF with gas fraction that we see is most prominent after the SP. This could be because at SP the central gas concentration may not be so strong, and so there is not intense AGN activity. However, SF after the third passage or FC is not significantly affected by gas fraction, perhaps because the AGN plays a stronger role, or maybe because more gas was consumed earlier.

\subsection{Black Hole Mass} \label{ssec:bh}

\begin{figure*}
\includegraphics[width=0.95\textwidth]{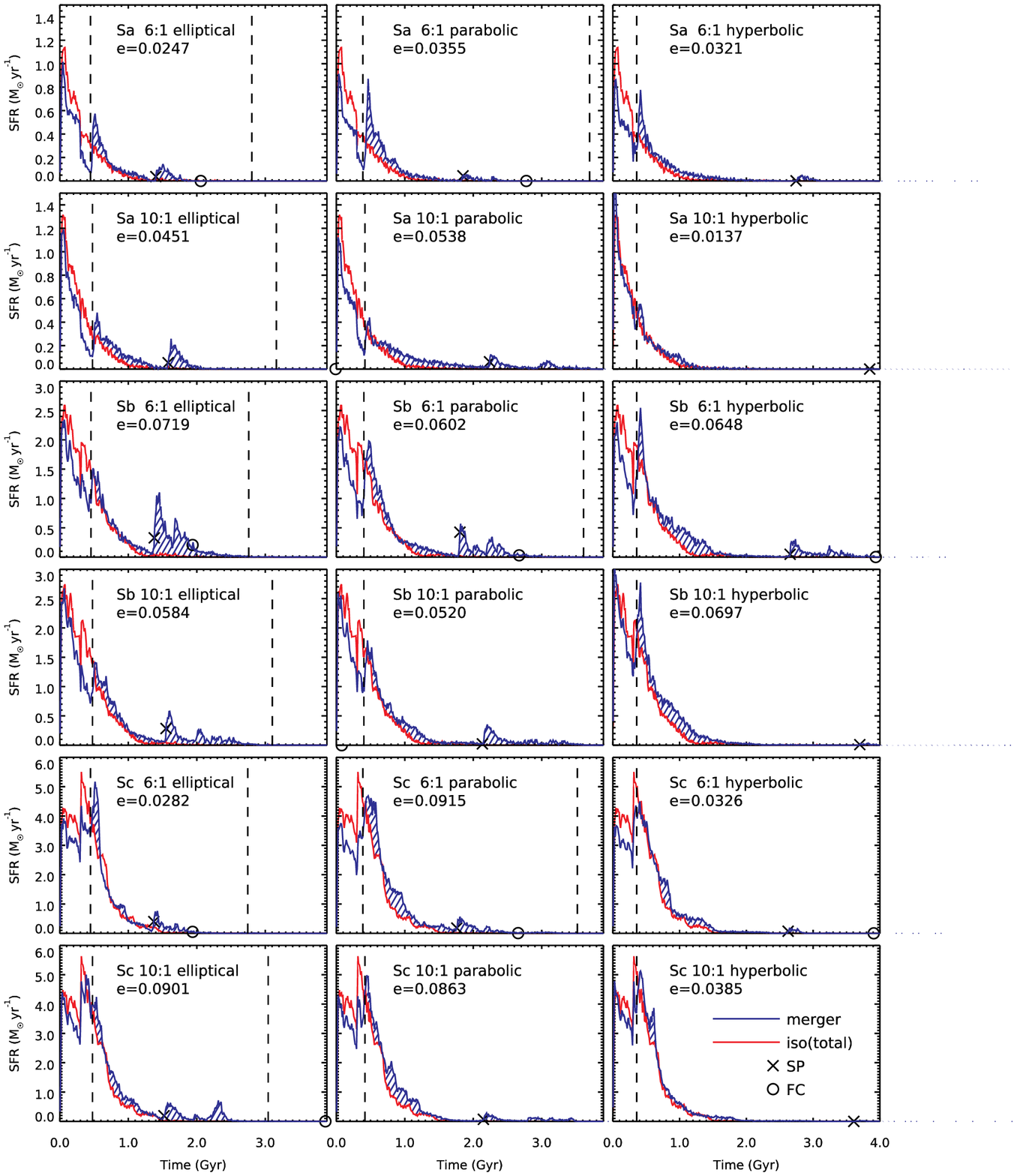}
\caption{SFHs and burst efficiencies of galaxies with different morphological types. The top two rows show the evolutions of Sa, the middle two rows show those of Sb and the bottom two show those of Sc type disk galaxies merging with secondary galaxies. The burst efficiencies are given for each combination of mass ratio and orbit.}
\label{fig:SFH_morphology}
\end{figure*}

Regarding the relation between SF and galactic morphology, earlier studies such as \citet{mh94} and C08 focused on the role of the bulge that prohibits gas inflow and starburst. However, the effect of SMBHs at the center of the bulge requires a thorough understanding given that the BH mass is coupled with the bulge mass or velocity dispersion \citep{metal98,mh03,getal09,betal11}.

To investigate the role of the BH mass, we use G1 as our primary galaxy, and vary the BH mass, while fixing every other parameter. Figure \ref{fig:SFH_BH} illustrates the results. We see no strong dependency on BH mass, except regarding the peak instantaneous SFR occurring at third passage or FC, where a more massive BH can slightly suppress the SF. But overall we find no clear trend between BH mass and burst efficiency.
 
\subsection{Morphology} \label{ssec:morphology}

Combining bulge fraction, gas fraction and BH mass, we produce model Sa- and Sc-like primary galaxies. Table \ref{tab:morp} summarizes these models. The Sb-like primary galaxy is G1, as mentioned in Section \ref{ssec:isoevo}. Generally, galaxy mass decreases from early to late spirals \citep{ketal15}, but we fix the stellar mass of the three primary galaxies for simplicity. We calculated the burst efficiencies of minor mergers of the combinations of three primary galaxies (Sa, Sb and Sc), secondary galaxies (G3 and G4) and orbits (elliptical, parabolic and hyperbolic).

Figure \ref{fig:SFH_morphology} presents these results. We do not see any clear trends with galaxy morphology, despite the fact that more late-type galaxies have smaller B/T and higher gas fraction. We believe this is because, as demonstrated in Section \ref{ssec:bulge} and Section \ref{ssec:gas}, the presence of AGN feedback has significantly weakened the dependency of the SFR on these parameters. 

% Table - Morphology
\begin{table}
  \begin{center}
    \caption{Summary of Primary Galaxies with Different Morphology.}
    \begin{tabular}{lccc}
    \hline \hline
    \multicolumn{1}{c}{ } & Sa & Sb (G1) & Sc \\
    \hline
    Stellar mass $(M_\odot)$ & $2.00 \times 10^{10}$ &  $2.00 \times 10^{10}$ 
                                            &  $2.00 \times 10^{10}$ \\
    Disk mass $(M_\odot)$ & $1.20 \times 10^{10}$ & $1.60 \times 10^{10}$ & $1.80 \times 10^{10}$ \\
    Bulge mass $(M_\odot)$ & $8.00 \times 10^9$ & $4.00 \times 10^9$ & $2.00 \times 10^9$ \\
    Gas mass $(M_\odot)$ & $1.33 \times 10^9$ & $4.00 \times 10^9$ & $6.00 \times 10^9$ \\
    $M_{\rm BH} (M_\odot)$ & $8.00 \times 10^6$ & $4.00 \times 10^6$ & $2.00 \times 10^6$ \\
    B/T & $0.40$ & $0.20$ & $0.10$ \\
    $f_{\rm gas}$ & $0.10$ & $0.20$ & $0.25$ \\
    $M_{\rm BH}/M_{\rm bulge}$ & 0.001 & 0.001 & 0.001 \\

    \hline
    \end{tabular}
    \label{tab:morp}
  \end{center}
\end{table}

\subsection{Parametrization in Semi-analytic Models} \label{ssec:sam}
 
The burst efficiency is parametrized as follows in certain semi-analytic galaxy formation models:
\begin{equation} \label{eq:burst}
e=e_{\rm 1:1}~(M_{\rm secondary}/M_{\rm primary})^\gamma
\end{equation}
Here, $e_{\rm 1:1}$ is the burst efficiency of an equal-mass merger. In the semi-analytical models of \citet{som08} and \citet{ly13}, this value is determined by the AGN simulations of \citet{retal06}. However, $\gamma$, which describes the relation between the galaxy mass ratio and merger-driven star formation, is determined by the non-AGN simulations of C08. Simulations by C08 yielded the value of $\gamma$ which depends on the bulge-to-total ratio of primary galaxies and \citet{som08} adopted this value in their model. However, as discussed in Section \ref{ssec:bulge}, the presence of AGN changes the relation between SF and bulge fraction. Therefore, we now use our simulations to obtain $\gamma$ considering the AGN feedback effect.

In our simulations, we derive different values of $\gamma$ for different B/T ratios. When the B/T ratios of the primary galaxies are 0.4, 0.2 and 0.0, the simulation results give values of $\gamma$ = 0.30, 0.38 and 0.42, respectively. The variation in $\gamma$ with B/T ratio is less than in the C08 simulations without AGN feedback ($\gamma =$ 0.61, 0.74 and 1.02 for B/T = 0.33, 0.17 and 0.00). However, readers should note that burst efficiency might be sensitive to the recipes of star formation and feedback in different kinds of simulations. Therefore, further investigations to understand the effect of AGN on $\gamma$ in the same conditions and the application to galaxy formation models are required in the future. The implication of the result in semi-analytic galaxy formation model is discussed in Section \ref{sec:disc}.

%% Section 4 - Mergers %%%%%%%%%%%%%%%%%%%%%%%%%%%%%%%%%%%%%%%%%%%%%%
%%%%%%%%%%%%%%%%%%%%%%%%%%%%%%%%%%%%%%%%%%%%%%%%%%%%%%%%%%%%%%%%%%%%%

%%%%%%%%%%%%%%%%%%%%%%%%%%%%%%%%%%%%%%%%%%%%%%%%%%%%%%%%%%%%%%%%%%%%%
%% Section 5 - Summary & Discussion %%%%%%%%%%%%%%%%%%%%%%%%%%%%%%%%%

\section{Summary \& Conclusions} \label{sec:disc}

Using the AMR hydrodynamics code {\sc ramses}, we have performed galaxy merger simulations to study the star formation of merging disk galaxies. We have run idealized merger simulations, changing different parameters such as the mass ratios of the two galaxies, the angle between the galactic spin plane and orbital plane of the secondary galaxy, type of orbit, primary galaxy's bulge fraction, gas fraction and black hole mass. With the burst efficiency previously introduced by C08, we quantify the merger-driven star formation of all merger simulations.

We perform approximately 70 individual numerical simulations of isolated and merging galaxies and obtain the following results.

\begin{enumerate}

\item
We find that in isolated galaxies, the presence of an AGN is not very significant for suppression of star formation. However, in merging galaxies the effect of the AGN is more prominent, as much more gas is funneled into the centers of the galaxies, feeding AGN activity. This result is consistent with \citet{nk13}. We additionally find that the timing is important. After first passage, the AGN does not play a strong role, but after final coalescence, it becomes highly significant in suppressing star formation.

\item
In our models, gas is efficiently funneled to the galaxy centers at the stage of the final coalescence. As a result, at this time AGN activity peaks and star formation is more efficiently suppressed. This is consistent with \citet{detal05}, however we consider a wide range of galaxy mass ratios, and by 10:1 we find that AGN activity enhancement is negligible. This is because minor mergers only weakly perturb the gas disk, as shown in \citet{cetal15}. We test the impact of having an AGN on the amount of stars produced during the simulations, and find the impact is stronger in major mergers.

\item
Coplanar mergers produce more new stars compared to mergers with other inclination angles. In our simulations, the retrograde merger shows a larger value of burst efficiency than that of the prograde merger because strong starburst occurs after first passage in the retrograde merger.

\item
No correlation between orbit eccentricity and star formation has been found. However the dependency on orbital parameters is complicated by the fact that more eccentric orbits tend to coalesce at later times, when there is less gas available for a star burst. Thus we find that the timing of the merger, and it relation to gas fraction, can play a dominant role in controlling the burst efficiency.

\item
Previously the bulge-to-total ratio was found to strongly effect the amount of stars produced in mergers. However, we find that with AGN feedback, merger-driven star formation becomes significantly less dependent on the bulge fraction of the primary galaxy. This is because the stronger concentration of gas in the galaxy centers that occurs in a bulgeless galaxy also triggers strong AGN activity, which in turn can suppress star formation.

\item
The gas fraction affects the SFR at first and second passage, but not as the galaxies approach final coalescence. This is because, at first and second passage, most of the star formation occurs away from the black hole. This result is in contrast to the simulations of C08, which could be the result of differing sub-grid treatment of star formation, or due to differing times when the merger occurs.

\item
We find that star formation rates are fairly independent of black hole mass, except during final coalescence when the gas is most funneled to the galaxy center.

\item
Sc type disk galaxies in our simulations do not always show the largest star formation and burst efficiency. This is because the AGN activity tends to suppress the additional star formation that would come from late-type galaxies being more gas rich, and less bulge dominated.

\item
We obtained burst efficiency fit with our simulations with AGN feedback. Compared to the non-AGN simulations by C08, our results suggest smaller dependence of $\gamma$ on bulge fraction.

\end{enumerate}

The inclusion of AGN feedback in merger simulations leads to a lower value of burst efficiency (Section \ref{ssec:sam}), hence, a decrease in the amount of merger-driven star formation is expected if the AGN effect is accurately considered in galaxy formation models. In our simulations, the total mass of stars formed by mergers could decrease by as much as a factor of two or three, in the presence of AGN feedback. However, we do not expect that this has a significant impact on the stellar masses of the global population of galaxies. Indeed, some semi-analytical models have demonstrated that merger-driven star formation only contributes a few percent of the total stellar mass in massive galaxies \citep{ly13}.

Our Sc type disk galaxies exhibit lower SFRs and burst efficiencies in some cases while those in observations (K14) show larger star formation enhancement by minor mergers. This is because star formation is suppressed efficiently by AGN feedback in our Sc models. However, as pointed out by \citet{nk13}, different AGN models yield different results for star formation in merging galaxies. Many studies on AGN focused on the calibration of model-related parameters using global properties such as the $M-\sigma$ relation in cosmological simulations \citep{bs09} and small scale processes around supermassive black holes were ignored. We expect that this leads to the exaggeration of the strength of AGN activity and adopting AGN prescriptions designed for cosmological simulations in galactic scale simulations causes the results to be inconsistent with observations. Some attempts at understanding AGN-related physics at smaller scales \citep{pr11,pr12} and making connections with the properties of host galaxies \citep{petal16} have been made, but there are still many gaps between AGN simulations at different physical scales. In that regard, further studies on small scale AGN physics and the development of AGN models applicable to galactic scale simulations are required.

%% Section 5 - Discussion %%%%%%%%%%%%%%%%%%%%%%%%%%%%%%%%%%%%%%%%%%%
%%%%%%%%%%%%%%%%%%%%%%%%%%%%%%%%%%%%%%%%%%%%%%%%%%%%%%%%%%%%%%%%%%%%%

\vspace{5cm}

%%%%%%%%%%%%%%%%%%%%%%%%%%%%%%%%%%%%%%%%%%%%%%%%%%%%%%%%%%%%%%%%%%%%%
%% Acknowledgement %%%%%%%%%%%%%%%%%%%%%%%%%%%%%%%%%%%%%%%%%%%%%%%%%%

\section*{}
We thank Jared Gabor for sharing his {\sc ramses} patch with us. J.P. thanks Paula Calderon Castillo for giving a lot of comments. 
S.K.Y. acknowledges support from the Korean National Research​ ​Foundation (NRF-2017R1A2A1A05001116). This study was performed under the umbrella of the joint collaboration between Yonsei University Observatory and the Korean Astronomy and Space Science Institute. Simulations were performed with KISTI supercomputer (program KSC-2015-C3-050). 

%% Acknowledgement %%%%%%%%%%%%%%%%%%%%%%%%%%%%%%%%%%%%%%%%%%%%%%%%%%
%%%%%%%%%%%%%%%%%%%%%%%%%%%%%%%%%%%%%%%%%%%%%%%%%%%%%%%%%%%%%%%%%%%%%

\vspace{10mm}

%%%%%%%%%%%%%%%%%%%%%%%%%%%%%%%%%%%%%%%%%%%%%%%%%%%%%%%%%%%%%%%%%%%%%
%% =============================================================== %%
%% end of text =================================================== %%
%% =============================================================== %%
%%%%%%%%%%%%%%%%%%%%%%%%%%%%%%%%%%%%%%%%%%%%%%%%%%%%%%%%%%%%%%%%%%%%%

%% references

\end{document}